\documentclass[12pt, a4paper]{article}
    \usepackage{tgpagella}
    \usepackage{mathpazo}
    \usepackage{enumitem}   
\usepackage{mathtools,amsthm,amssymb}
    \usepackage{amsmath}
    \usepackage{pgfplots}
    \usepackage{natbib}
    \usepackage{eufrak}
    \usepackage[normalem]{ulem}
        \setcitestyle{open={(},close={)}}
    \usepackage{appendix}
    \usepackage{pgfplots}
        \pgfplotsset{width=12cm,compat=1.9}
        \usepackage{pst-func}
        \psset{unit=2cm}
        \usepgfplotslibrary{fillbetween}
    \usepackage{hyperref}
    \usetikzlibrary{patterns, arrows.meta}
        \hypersetup{colorlinks=true,urlcolor=blue,citecolor=blue,linkcolor=red}
    \usepackage[center]{caption}
    \usepackage{subcaption}
	\usepackage{comment}
	 \usepackage[onehalfspacing]{setspace}
	\usepackage{bigints}
	\usepackage{booktabs}
	\usepackage{epsfig}
	\usepackage{graphics}
	\usepackage{lscape}
	\usepackage[english]{babel}
	\usepackage{color}
	\usepackage{marvosym}
	\usepackage[margin=1in]{geometry}
	\usepackage{tikz}
	\usepackage{chngcntr}
    \usepackage{apptools}
    \usepackage{tikz, smartdiagram}
     \usetikzlibrary{shapes.multipart}  
\usetikzlibrary{decorations.pathreplacing,calligraphy}
     \tikzset{
         state/.style={rectangle split, draw=black, text width=3cm}
     }
     \tikzset{every picture/.append style={remember picture}}

    \AtAppendix{\counterwithin{assumption}{section}}
	
	\DeclareMathOperator*{\argmax}{arg\,max}
	\DeclareMathOperator*{\argmin}{arg\,min}

    \newcommand{\bp}{\mathbf{p}} 
     
    \newcommand{\bq}{\mathbf{q}} 
    \newcommand{\bh}{\mathbf{h}} 
    \newcommand{\bM}{\mathbf{M}} 
     
    \newcommand{\bP}{\mathbf{P}}

    \newcommand{\Cov}{\text{Cov}} 
     
    \newcommand{\myvec}{\text{vec}} 
        
    \newtheoremstyle{indented}{20pt}{20pt}{\addtolength{\leftskip}{0.0em}}{}{\bfseries}{.}{.5em}{}
    \newtheoremstyle{indented_nonbold}{15pt}{15pt}{\addtolength{\leftskip}{1.5em}}{}{\itshape}{.}{.5em}{}

    \theoremstyle{indented}
    \theoremstyle{indented}\newtheorem {prop}{Proposition}
    \theoremstyle{plain}\newtheorem {thm}{Theorem}
    \newtheorem*{thm*}{Theorem}
    \theoremstyle{indented}
    \theoremstyle{indented}
    \theoremstyle{indented}
    \theoremstyle{indented}\newtheorem {assumption}{Assumption}
    \theoremstyle{indented}\newtheorem{mydef}{Definition}
    \theoremstyle{indented}
    \theoremstyle{indented} 
    \theoremstyle{remark} 
    \theoremstyle{indented} \newtheorem{rem}{Remark}
    \theoremstyle{remark} 

    \title{Beyond the Mean: Testing Consumer Rationality through Higher Moments of Demand}
    
    \author{Sebastiaan Maes\thanks{University of Antwerp; \href{mailto:sebastiaan.maes@uantwerpen.be} {\tt sebastiaan.maes@uantwerpen.be}}\quad  Raghav Malhotra\thanks{University of Leicester; \href{mailto:r.malhotra@leicester.ac.uk} {\tt r.malhotra@leicester.ac.uk}}
    \footnote{Sebastiaan thanks Andr\'e Decoster for his supervision. Raghav thanks Herakles Polemarchakis, Costas Cavounidis, and Robert Akerlof for their supervision. We are grateful to Debopam Bhattacharya, Pablo Becker, Luis Candelaria, Laurens Cherchye, Daniele Condorelli, Sam Cosaert, Ian Crawford, Liebrecht De Sadeleer, Geert Dhaene, Peter Hammond, Yuichi Kitamura, Kenichi Nagasawa, Eric Renault, Camilla Roncoroni, and Ao Wang for fruitful discussions. We also thank conference and workshop participants at Leuven, Tilburg, Warwick, AEE, ESEM, ECORES, ESCOE, EWMES and LAGV. Sebastiaan benefited from doctoral and postdoctoral fellowships of the Research Foundation Flanders (grants 11F8919N and 12C8623N). Part of this research has been conducted during a visit to CRETA at Warwick. The results and their interpretation are the authors' sole responsibilities.}}
    
    \date{\today\bigskip}

\begin{document}	
    \begin{titlepage}
        \maketitle
        \thispagestyle{empty}
          \begin{abstract}
           We study a setting where an analyst has access to purely aggregate information about the consumption choices of a heterogenous population of individuals. We show that observing the statistical moments of market demand allows the analyst to test aggregate data for rationality. Interestingly, just the mean and variance of demand carry observable restrictions. This is in stark contrast to impossibility result of the Sonnenschein-Mantel-Debreu theorem, which shows that aggregate demand carries no observable restrictions at all. We leverage our approach to deliver a characterization of rationality in terms of moments for the common two-good case. We illustrate the usefulness of moment-based restrictions through two applications: (i) improving the precision of demand and welfare estimates; and (ii) testing for the existence of a welfare-relevant representative consumer.
        \end{abstract}
        
    
    \end{titlepage}

\section{Introduction}
Does individual utility maximization generate observable restrictions on aggregate variables? Work in the early 1970s showed that when individuals in a population differ arbitrarily, any function could manifest as an average of individually rational demand functions.\footnote{For an comprehensive reviews, see \citet*{RizviSMDreview, chiappori2011new}.} These findings led to widespread skepticism about the ability to test rational behavior theories and apply consumer theory to aggregate data. \citet*{arrow1990economic}, for instance,  stated that \textit{``in the aggregate, the hypothesis of rational behaviour has, in general, no
implications"} and that \textit{"if agents are different in unspecifiable
ways, then [...] very few, if any, inferences can be made"}.

Contrary to this widespread pessimism, our paper suggests that the difficulty in testing theories of rational behavior mainly stems from an over-reliance on analyzing average demand alone. We demonstrate that rationality does yield testable restrictions when the analyst observes slightly more. Specifically, the restrictions rationality imposes on demand data grow with the number of observable statistical moments. Even just the mean and variance of demand deliver a test of rationality. Our analysis therefore offers a novel and nuanced perspective that underscores the fragility of previous negative results. Additionally, establishing moment inequalities for demand models, this paper connects the extensive literature on partial identification through moment inequalities with nonparametric analysis of demand models.

We characterize rationality in terms of moment sequences for the two-good case. We then present the maximal test of rationality based on moment data and operationalize it using a simple algorithm. A numerical example shows that in practice, the first few moments can capture a large amount of irrationality in a population.  

If there are more than two goods, we demonstrate that the population moments do not allow the analyst to test individuals for symmetry of the Slutsky matrix but only negative semidefiniteness. In a sense, this corresponds to preferences which obey the weak axiom of revealed preference but not the strong axiom. However, negative semidefiniteness has observational implications.\footnote{\citet{fosgerau2023nontransitive} prove a similar claim for discrete choice.}
However, if the analyst assumes that the population obeys symmetry, the average substitution matrix can be point identified from the first two moments.

Before proceeding to formal results, we briefly summarize the history of the literature. \citet*{sonnenschein73} proved that with two goods, Walras' law and homogeneity of degree zero are the only restrictions on excess demand. \citet*{debreu74} and \citet*{mantel74,mantel76} extended the argument to general economies. These negative results are collectively known as the Sonnenschein-Mantel-Debreu (SMD) theorem.\footnote{This result had a significant and resounding impact on economic theory. James Tobin, who firmly held that economics can and should alleviate need and improve general welfare, considered the
SMD theorem as a result that should not have been proved \citep{polemarchakis2004}.} However, more recent work has shown that this view is overly pessimistic. \citet*{brownmatzkin96}, \citet*{chiapporiekeland02,chiapporiekeland04}, \citet*{kubler03}, and \citet*{chiapporietal04} show that general equilibrium theory does generate robust predictions, but only with access to individual-level data. For market demand, \citet*{ChiapporiEkeland1999} show that any analytic market demand function satisfying homogeneity and Walras' law can locally be decomposed into finitely many rational demand functions. This paper goes one step further, highlighting that even individual data is unnecessary. Just a few extra features of the aggregate distribution of consumption yield sharp predictions of utility maximization.\footnote{\citet*{hildenbrand83, hildenbrand94} and \citet*{hardle1991empirical} impose restrictions on the variance such that aggregate demand obeys the so-called \emph{Law of Demand}. Instead, we tackle the inverse problem.} 

We illustrate the usefulness of our restrictions through two applications. Firstly, we show how these restrictions can be leveraged to improve the precision of the estimated moments of demand.\footnote{In a related paper, \cite{maesmalhotra2023robust} show that these moments delivers best approximations to welfare changes using cross-sectional data, and can be used for other applications like  the estimation of price indices and the elasticity of taxable income.} We advance an empirical Bayes approach in the spirit of \citet{fesslerkasy} to shrink the unconstrained estimates towards the theoretical restrictions, if the latter are not rejected by the data. This allows taking advantage of the information in the theoretical restrictions while mitigating the impact of potential misspecification. Secondly, we devise a test for the existence of a welfare relevant consumer. It is well known that a \emph{positive} representative agent may not be welfare-relevant; he may even be Pareto inconsistent, i.e., preferring one situation to another even though all agents in society prefer the reverse \citep*{dow1988consistency,jerison1984social}. We construct a test to assess the existence of a \emph{normative} representative agent.
 
Our results relate to the literature on stochastic rationalizability where the entire cross-sectional distribution of demand can be observed. In the many-good case, \citet*{Hoderleinstoye2014,Hoderleinstoye2015} and \citet*{Dettehoderlein2016} derive and test restrictions on the quantiles of demand. We highlight the theoretical connection between our moments-based approach and the quantile-based approach on microdata.\footnote{\citet*{Dettehoderlein2016} and \citet*{hausman2016individual} show that with two goods, a population is rational if and only if its conditional quantile functions satisfy the Slutsky restrictions.} We show that our approach implicitly weights the restrictions provided by the quantile-based approach. In a related exercise, \cite{hoderlein2011many} uses techniques similar to ours to bound the proportion of individuals in a population who could satisfy rationality. \citet*{kitamurastoye18} provide tests based on revealed preference inequalities for finitely many demand distributions at different prices.\footnote{By contrast, our results assume differentiable demands but are valid at the population level.}
An advantage of our approach is that it can also be employed when researchers do not observe the entire demand distribution but only some coarse moments. Moreover, our moment-based results scale naturally to the many-good case, whereas the quantile-based approach does not.

\section{Conceptual framework}\label{sec:setup}

Our conceptual framework allows for unrestricted, unobserved heterogeneity in preferences. For ease of exposition, we suppress all observed individual characteristics; all results in this paper can be thought of as conditional on these covariates.

\subsection{Consumer demand}
We consider the standard model of utility maximization under a linear budget constraint. Let $\Omega$ denote the universe of preference types. Every preference type $\omega \in \Omega$ can be considered an individual with preferences over bundles of $(k+1)$ goods $\bq$. We assume the set of bundles is compact and convex and denote it as $\mathcal{Q} \subseteq \mathbb{R}_{++}^{k+1}$. Preferences are assumed to be representable by smooth, strictly quasi-concave utility functions $u^\omega : \mathcal{Q} \to \mathbb{R}$. Prices are denoted $\bp \in \mathcal{P}\subset \mathbb{R}_{++}^{k+1}$ and income, $y \in \mathcal{Y} \subset \mathbb{R}_{++}$. We call a pair $(\bp, y)$ a budget. 

Individual demand functions $\bq^\omega(\bp, y) : \mathcal{P} \times \mathcal{Y} \to \mathcal{Q}$ arise from individuals maximizing their utility subject to a linear budget constraint,
\begin{equation*}
    \bq^\omega(\bp,y)=\argmax_{\bp\cdot\mathbf{q}\leq y : \mathbf{q} \in \mathcal{Q}}u^\omega(\bq).
\end{equation*}
Every uncompensated demand function $\bq^\omega$, generates a compensated demand function $\bh^\omega(\bp, u) : \mathcal{P} \times \mathbb{R} \to \mathcal{Q}$ defined as 
\begin{equation*}
    \bh^\omega(\bp,u)=\argmin_{\mathbf{q} \in \mathcal{Q}}\{\bp\cdot \mathbf{\bq} | u^\omega(\bq)\geq u\}.
\end{equation*}
The Slutsky equation
\begin{equation}\label{eq:slutsky}
    \frac{\partial}{\partial \bp} \bq^\omega(\bp,y) = \frac{\partial}{\partial \bp} \bh^\omega(\bp, u) -  \frac{\partial}{\partial y}\bq^\omega(\bp,y) \bq^\omega(\bp,y)^\intercal,
\end{equation}
links both demand functions. We will omit the demand and price for the $(k+1)$st good using Walras' law.

\subsection{Moments of demand}
In the two-good case, integrating out unobserved preference heterogeneity, we can express the $n$th (non-central) \emph{conditional moment of demand} as
\begin{equation}\label{eq:condmonsing}
    \begin{split}
        M_n(p,y)&=     \mathbb{E}[q^\omega(p,y)^n \mid p,y]\\
        &=\int q^\omega(p,y)^n dF(\omega), 
    \end{split}
\end{equation}
where $F(\omega)$ denotes the distribution of preference types.\footnote{These moments can be readily inferred from cross-sectional data if preference types are distributed independently of prices and income \citep*{hausman2016individual, blomquistnewey}. Under budget set endogeneity, a control function approach can be used \citep{hoderlein2011many}.} By Walras' law, it suffices to consider scalar demand, thus these only depend on the price of the modelled good. 
 

In the many-good case, one can express the $n$th conditional moment of demand by means of the symmetric $n$ tensor $\mathbf{T}^\omega_n(\bp,y)$ for which the element $t^\omega_{i_1, i_2, \dots i_n}(\bp,y) = q^\omega_{i_1}(\bp,y)q^\omega_{i_2}(\bp,y)\dots q^\omega_{i_n}(\bp,y)$ with $i_1, i_2, \dots, i_n \in \{1, 2, \dots, k\}$. We define the generalized \emph{tensor form} of $\mathbf{T}^\omega_n(\bp,y)$ with respect a vector $\mathbf{v} \in \mathbb{R}^{k}$ as the multilinear function 
\begin{equation*}
    \begin{split}
        \mathbf{v}(**)\mathbf{T}^\omega_n(\bp,y) &=\mathbf{T}^\omega_n(\bp,y)(\underbrace{\mathbf{v}\times \mathbf{v} \dots \times \mathbf{v}}_\text{$n$ times})\\
        &= \sum_{i_1, i_2, \dots, i_n = 1}^{k} t^\omega_{i_1, i_2, \dots, i_n}(\bp,y) v_{i_1}v_{i_2}\dots v_{i_n}.
    \end{split}
\end{equation*}
Again, by integrating out unobserved preference heterogeneity, we can express the $n$th (non-central) conditional moment of demand as
\begin{equation}\label{eq:condmom}
    \begin{split}
        \bM_n(\bp,y)&= \mathbb{E}\left[\mathbf{T}^\omega_n(\bp,y) \right]\\
        &=\int\mathbf{T}^\omega_n(\bp,y)dF(\omega \mid \bp,y). \\
    \end{split}
\end{equation}

Henceforth expectations are always conditional on a budget $(\bp,y)$: i.e., for a random variable $z(\bp,y)$, we will write $\mathbb{E}[z(\bp,y)] = \mathbb{E}[z(\bp,y) \mid \bp,y]$. For the sake of brevity, we will call these \emph{conditional moments} simply \emph{moments}. We define a \emph{moment sequence} as the (possibly infinite) sequence $\{\bM_i(\bp,y)\}_{i=1}^n$ of the first $n$ moments of demand.
\subsection{Rationalizability}
Let $\{\mathbf{a}_i(\bp,y)\}_{i=1}^r$  be a sequence where each $\mathbf{a}_i(\bp,y):\mathcal{P} \times \mathcal{Y} \to \mathbb{R}^{(k-1)^i}$ is function which maps budget sets to tensor forms of (weakly) increasing dimension. We say $\{\mathbf{a}_i(\bp,y)\}_{i=1}^k$ is \emph{rationalizable} \textit{around $(\bp_0,y_0)$} if there exists a universe of preference types $\overline{\Omega}$ and a probability measure $\overline{F}(\omega)$ over these types such that
\begin{equation*}
    \mathbf{a}_i(\bp,y)=\int\overline{\mathbf{T}}^\omega_i(\bp,y) d\overline{F}(\omega), \quad \forall i \leq k,
\end{equation*}
holds for an open set around the budget set $(\bp_0,y_0)$, and $\overline{\mathbf{T}}^\omega_i$ is generated by a rational demand function $\overline{\bq}^\omega$ for all $\omega \in \Omega$. A demand function is called \emph{rational} when it obeys Slutsky symmetry and negative semidefiniteness (NSD), homogeneity of degree zero, and Walras' law.

Technical conditions are relegated to Appendix~\ref{app:techcon}. In particular, we assume that the conditions for the dominated convergence theorem hold such that derivative and integral operators can be interchanged.


\section{Main results}\label{sec:momrational}
We start with the two-good case in Section~\ref{sec:twogood} and then move to the many-good case in Section~\ref{sec:manygood}. 

\subsection{The two-good case}\label{sec:twogood}
\begin{prop}\label{prop:res2mom}
For all budgets $(p,y) \in \mathcal{P} \times \mathcal{Y}$, it holds that 
\[\frac{1}{n+1} \frac{\partial}{\partial p} M_{n+1}(p,y) + \frac{1}{n+2}  \frac{\partial}{\partial y}M_{n+2}(p,y)\leq 0, \qquad  n \in \mathbb{N}.\]

\end{prop}
\begin{proof}
    By the Slutsky equation~\eqref{eq:slutsky}, for all types $\omega \in \Omega$ at all budget sets $(p,y) \in \mathcal{P} \times \mathcal{Y}$, it holds that
\begin{equation*}
    0 \geq  \frac{\partial}{\partial p} q^\omega(p,y) + q^\omega(p,y)  \frac{\partial}{\partial y}q^\omega(p,y). 
\end{equation*}
Since demand is positive, multiplying this expression by $q^\omega(p,y)^n$ preserves this inequality.
\begin{equation*}
    \begin{split}
       0 &\geq q^\omega(p,y)^{n} \left[\frac{\partial}{\partial p} q^\omega(p,y) + q^\omega(p,y)  \frac{\partial}{\partial y}q^\omega(p,y)\right]\\
        &= \frac{1}{n+1} \frac{\partial}{\partial p} q^\omega(p,y)^{n+1}+ \frac{1}{n+2}  \frac{\partial}{\partial y} q^\omega(p,y)^{n+2},
    \end{split}
\end{equation*}
where the equality follows from the product role for derivatives. Taking expectations at both sides gives
\begin{equation*}
    \begin{split}
        &=\int \left[\frac{1}{n+1} \frac{\partial}{\partial p} q^\omega(p,y)^{n+1}+ \frac{1}{n+2}  \frac{\partial}{\partial y} q^\omega(p,y)^{n+2}\right] dF(\omega)\\
        &=\frac{1}{n+1} \frac{\partial}{\partial p} M_{n+1}(p,y) + \frac{1}{n+2}  \frac{\partial}{\partial y}M_{n+2}(p,y),
    \end{split}
\end{equation*}
the last equality following from the interchange of integration and differentiation.
\end{proof}
Nevertheless, the inequalities in Proposition~\ref{prop:res2mom} turn out not to be {exhaustive} restrictions which moments provide. To obtain the exhaustive effects of rationality on moments, we define objects called polynomial translations. Let the $n$th monomial translation be
\begin{equation*}
    \begin{split}
        \Gamma^\omega_n(p,y)
        &=q^\omega(p,y)^{n} \left[\frac{\partial}{\partial p} q^\omega(p,y) + q^\omega(p,y)  \frac{\partial}{\partial y}q^\omega(p,y)\right],
    \end{split}
\end{equation*}
and denote its expectation as
\begin{equation*}
    \Gamma_n(p,y) = \mathbb{E}\left[\Gamma^\omega_n(p,y)\right].
\end{equation*}
Since $\Gamma_n(p,y)=\frac{1}{n+1} \frac{\partial}{\partial p} M_{n+1}(p,y) + \frac{1}{n+2}  \frac{\partial}{\partial y}M_{n+2}(p,y)$, we can rewrite Proposition~\ref{prop:res2mom} in terms of translations:
\[\text{For all budgets } (\bp,y) \text{ and } n \geq 0: \quad \Gamma^\omega_n(p,y) \leq 0. \]


 Let $\mathbb{Q}\left[\mathbb{R}\right]$ be the set of polynomials over the real numbers with rational coefficients $\{a_i\}$. For any polynomial $\pi_n^\omega(p,y) = \sum_{i=1}^n a_i (q^{\omega}(p,y))^n \in \mathbb{Q}\left[\mathbb{R}\right]$, we define its associated polynomial translation as:
\begin{equation*}
    \begin{split}
        \Lambda^\omega_{\pi_n}(p,y)
        &=\pi^\omega_n(p,y) \left[\frac{\partial}{\partial p} q^\omega(p,y) + q^\omega(p,y)  \frac{\partial}{\partial y}q^\omega(p,y)\right] \\
        &= \sum_{i=1}^n a_i (q^{\omega}(p,y))^n\left[\frac{\partial}{\partial p} q^\omega(p,y) + q^\omega(p,y)  \frac{\partial}{\partial y}q^\omega(p,y)\right] \\
        &=  \sum_{i=1}^n a_i \Gamma^\omega_n(p,y),
    \end{split}
\end{equation*}
and denote its expectation as
\begin{equation*}
    \begin{split}
        \Lambda_{\pi_n}(p,y) & =\mathbb{E}\left[\Lambda^\omega_{\pi_n}(p,y)\right] \\
        &= \mathbb{E}\left[\sum_{i=1}^n a_i \Gamma^\omega_n(p,y)\right] \\
        &= \sum_{i=1}^n a_i \Gamma_n(p,y),\\
    \end{split}
\end{equation*}
where the last equality follows from the linearity of the expectation operator. We now give exhaustive restrictions in terms of moments of demand.



\begin{thm}\label{thm:rationalizability}
    In the two-good case, the following statements are equivalent:
    \begin{enumerate}
        \item A demand distribution can be generated by a rational population.
        \item For any polynomial $\pi_n(p,y)$ that is positive in the support of the distribution of demand at $(p,y)$, it holds that $\Lambda_{\pi_n}(p,y) \leq 0$.
    \end{enumerate}
\end{thm}

\begin{proof}
The $(1)\Longrightarrow(2)$ part simply follows from any polynomial transformation being a sum of monomial transformations, thus requiring negativity. For the $(2)\Longrightarrow(1)$ part, we proceed by means of proof by contradiction. \citet*{hausman2016individual} show that negativity of the quantile demand function characterizes rationalizability. Suppose $(2)$ holds, but negativity is contradicted at some quantile. This would imply that there is some quantile $\tau \in (0,1)$, and some quantile demand $Q({\tau} \mid p,y) = \inf\{q : \Pr[q^\omega(p,y) \leq q \mid p,y] \geq {\tau}\}$ such that 
\begin{equation*}
    \frac{\partial}{\partial p} Q({\tau}\mid p,y) + Q({\tau} \mid p,y)  \frac{\partial}{\partial y}Q({\tau} \mid p,y)>0. 
\end{equation*}
We can pick a sequence of polynomials $\{\pi_n\}_{n=1}^\infty$ such that
\begin{equation*}
    \lim_{n\to \infty}\{\pi_n\}\to \delta({Q({\tau}|p,y)}),
\end{equation*}
where $\delta$ is the Dirac delta function.\footnote{To be precise, one should pick a set of sequences of polynomials that uniformly converge in a neighborhood of the budget $(p,y)$. Therefore, derivatives with respect to $p$ and $y$ are well-defined.} Therefore, by continuity of $\Lambda_{\pi_n}$, we have that
\begin{equation*}
    \lim_{n\to \infty}\{\Lambda_{\pi_n}( p,y)\}\to\left[\frac{\partial}{\partial p} Q({\tau}\mid p,y) + Q({\tau} \mid p,y)\frac{\partial}{\partial y}Q({\tau} \mid p,y)\right]>0,
\end{equation*}
which means that beyond some finite $n \in \mathbb{N}$, negativity must be contradicted. This would in turn contradict $(2)$, hence proving the theorem.
\end{proof}

\paragraph{Testing rationality.}
The equivalence in Theorem~\ref{thm:rationalizability} can be used to construct a semi-decidable test.\footnote{This test has the property that no rationalizable distribution is every rejected and all non-rationalizable distributions are eventually rejected.} Let $\mathbb{Q}_+\left[\mathbb{R}\right] = \{\pi \in \mathbb{Q}\left[\mathbb{R}\right] \mid x \in [0, y/p] \implies \pi(x)\geq 0\}$ be the set of polynomials over the real numbers with rational coefficients that are positive for $x \geq 0$. Since the rational numbers are countable, so is the set $\mathbb{Q}_+\left[\mathbb{R}\right]$; one can therefore pick an enumeration $\{\pi_n\}_{n=1}^\infty$ of this set. A simple semi-decidable test would consist of the following iterative scheme at step $n$:
    \begin{enumerate}
    \item If $\Lambda_{\pi_n}(p,y) \leq 0$, move to the $(n+1)$st step.
    \item If $\Lambda_{\pi_n}(p,y) > 0$, stop and reject the distribution.
\end{enumerate}
    
    The first part follows directly from Theorem~\ref{thm:rationalizability}. The second part follows from the fact that if the distribution is not rationalizable, there exists some polynomial $\pi$ which has a positive translation. Since $\{\pi_n\}_{n=1}^\infty$ is countable, there must exist some $n$ where $\pi_n$ has a positive translation, leading to rejection.

\begin{rem}\label{rem:test2}
    In the case where only the zeroth and first monomial translation can be computed (or equivalently, the first three moments can be observed), only linear polynomials enter the analysis, which makes testing much simpler. Denote the support of demand at budget $(p,y)$ as $0\leq q_{min} \leq q_{max}\leq y/p$. In terms of the first two translations, only four polynomials need to be checked for negativity: (i) $1$; (ii) $x$; (iii) $-q_{min}+ x$; and (iv) $q_{max}- x$. This translates to the conditions:
    \begin{align*}
    \Gamma_0(p,y)\leq 0,\\
    \Gamma_1(p,y)\leq 0,\\
    -q_{min}\Gamma_0(p,y)+  \Gamma_1(p,y)\leq 0,\\
    q_{max}\Gamma_0(p,y)-     \Gamma_1(p,y) \leq 0.
\end{align*}
    This means that in additional to monomial negativity, only $q_{max}\Gamma_0(p,y)\leq \Gamma_1(p,y) \leq q_{min}\Gamma_0(p,y)$ needs to be checked. Figure~\ref{fig:semitest} shows the admissible set of solutions shaded in red. 
    
    \begin{figure}[h!]
    \centering    
 \begin{tikzpicture}
   \begin{axis}[
   		xmin=-3, xmax=3,
   		ymin=-10, ymax=10,
   		xtick distance=1, ytick distance=4 ]

    \addplot [domain=-3:3, samples=100, name path=f, thick, color=red!50]
        {2*x};

    \addplot [domain=-3:3, samples=100, name path=g, thick, color=blue!50]
        {5*x};

      \draw [dashed, opacity=0.4] (axis cs:{0,10}) -- (axis cs:{0,-10});
      \draw [dashed, opacity=0.4] (axis cs:{3,0}) -- (axis cs:{-3,0});
      \addplot[red, opacity=0.7, pattern= north west lines,pattern color=red] fill between[of=f and g, soft clip={domain=-3:0}];

      \node[color=red, font=\footnotesize] at (axis cs: -1.6,-0.4) {$q_{min}\Gamma_0( \bp,y)= \Gamma_1(\bp,y)$};
      \node[color=blue, font=\footnotesize] at (axis cs: -0.05,6) {$q_{max}\Gamma_0( \bp,y)=\Gamma_1(\bp,y)$};
    \end{axis}
  \end{tikzpicture}
      \caption{Test for rationality based on the three first moments of demand}
    \label{fig:semitest}
\end{figure}
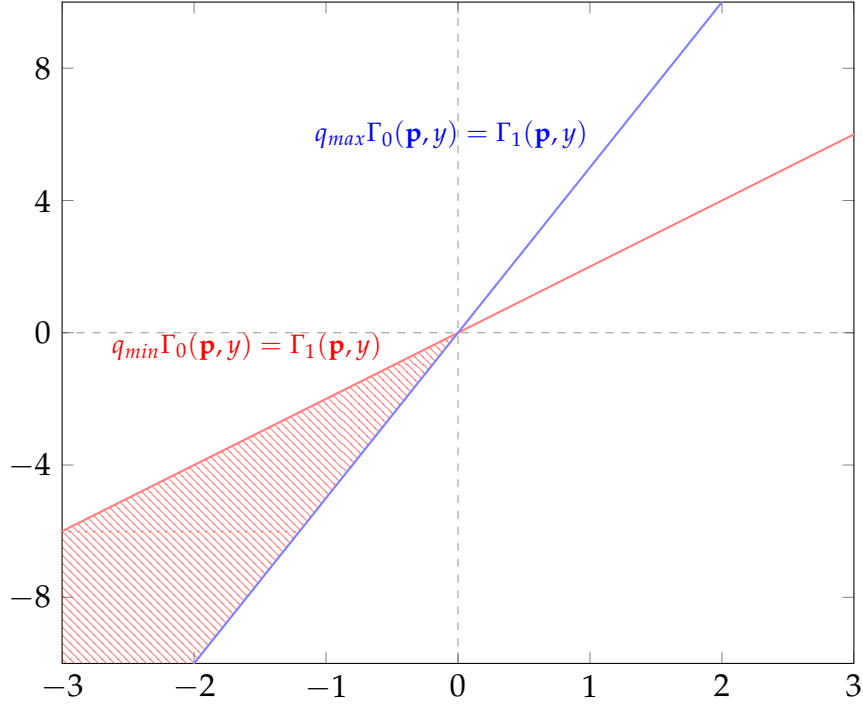
    
\end{rem}

\begin{rem}
    This can procedure can be further. Let the support of demand at budget $(p,y)$ be $(0,y/p]$. Without loss of generality this can be rescaled into $(0,1]$. Let $b_{\nu,n}(q) = \binom{n}{\nu} q^\nu (1-q)^\nu, v = 0, \dots, n$ be the $n+1$ Bernstein polynomials of degree $n$. These are positive on the whole domain $[0,1]$. They form a basis for the polynomials of degree $\leq n$ with real coefficients: $\pi_n(q) = \sum_{\nu = 0}^n \beta_\nu b_{\nu, n} (q)$, where $\beta_\nu \in \mathbb{R}$ are Bernstein coefficients. This implies the following test procedure for an $n$th degree polynomial (on the basis of the first $n+2$ moments). The test for polynomials of degree $\leq n$ is equivalent to check for all
    \begin{equation*}
        \Big\{\beta_\nu : \pi_n(q) = \sum_{\nu = 0}^n \beta_\nu b_{\nu, n} (q) \geq 0, \forall q \in [0,1] \Big\},
    \end{equation*}
    whether $\sum_{\nu = 0}^n \beta_\nu \Lambda_{b_{\nu, n}}( p,y)\leq 0$.
It is immediate that Remark~\ref{rem:test2} is a special case, as the zeroth and first degree basis functions of Bernstein polynomials are $b_{00}(q) = 1$, $b_{01}(q) = 1-q$, and $b_{11}(q) = q$.

A practical implementation is as follows. Construct a grid over $(0,1]$ and check for every grid point $q'$ the linear program
\begin{equation*}
    \begin{split}
        \max_{\boldsymbol{\beta}} \sum_{\nu = 0}^n \beta_\nu \Lambda_{b_{\nu, n}}( p,y) \text{ subject to } \sum_{\nu = 0}^n \beta_\nu b_{\nu, n} (q') \geq 0.
    \end{split}
\end{equation*}
If this maximand is positive at at least one grid point, rationality is rejected.
\end{rem}

\subsection{The many-good case}\label{sec:manygood}
\paragraph{The role of Slutsky symmetry.} 
We now consider the case where we have many goods. The main difference with the two-good case is that Slutsky symmetry becomes a material restriction with many goods. 

\begin{prop}\label{prop:symmetry}
Without Slutsky symmetry being imposed, $\mathbb{E}\left[\frac{\partial}{\partial y}\bq^\omega(\bp,y) \bq^\omega(\bp,y)^\intercal\right]$ is not identified from the first two moments of demand.
\end{prop}
\begin{proof}
For simplicity, we consider the case with three goods, with the third being the numeraire. 
From the definition of the second moments, it follows that
\begin{equation*}
    \bM_2(\bp,y) = 
   \mathbb{E} \begin{bmatrix}
   (q_{1}^\omega(\bp,y))^2 && q_{1}^\omega(\bp,y)q_{2}^\omega(\bp,y)\\
     q_{1}^\omega(\bp,y)q_{2}^\omega(\bp,y) &&  (q_{2}^\omega(\bp,y))^2
    \end{bmatrix},
\end{equation*}
which is a symmetric matrix. However, one needs to identify
\begin{equation*}
    \mathbb{E} \begin{bmatrix}
        q_{1}^\omega(\bp,y) \frac{\partial}{\partial y}q_{1}^\omega(\bp,y) & q_{1}^\omega(\bp,y)  \frac{\partial}{\partial y}q_{2}^\omega(\bp,y) \\
        q_{2}^\omega(\bp,y) \frac{\partial}{\partial y}q_{1}^\omega(\bp,y) & q_{2}^\omega(\bp,y)  \frac{\partial}{\partial y}q_{2}^\omega(\bp,y)
    \end{bmatrix}.
\end{equation*}
Even though the diagonal terms of this matrix are pinned down (through the derivative with respect to $y$), the off-diagonal terms cannot be identified because the information in the variance is redundant. In particular, we can identify 
\begin{equation*}
    \mathbb{E}\left[\frac{\partial}{\partial y}(q_{1}^\omega(\bp,y)q_{2}^\omega(\bp,y))\right] = \mathbb{E}\left[q_{1}^\omega(\bp,y) \frac{\partial}{\partial y}q_{2}^\omega(\bp,y) +  \frac{\partial}{\partial y}q_{1}^\omega(\bp,y)q_{2}^\omega(\bp,y)\right],
\end{equation*}
but not each term at the right-hand side separately. This means there can exist different models that disagree on the value of $\mathbf{E}\left[ \frac{\partial}{\partial y}\bq^\omega(\bp,y) (\bq^\omega(\bp,y))^\intercal\right]$ but are observationally equivalent in terms of the first two moments of demand.
\end{proof}

Without imposing Slutsky symmetry, the matrix $\mathbb{E}\left[ \frac{\partial}{\partial y}\bq^\omega(\bp,y) \bq^\omega(\bp,y)^\intercal\right]$ that captures income effects is not automatically identified from the first two conditional moments of demand. This is due to the fact that the variance of demand being symmetric imposes a loss of ``degrees of freedom''. This is different from the two-good case, where there is no loss of information because symmetry holds trivially.

Proposition~\ref{prop:symmetry} shows that if we remain agnostic about rationality, income effects are not identified from the first two moments of demand. However, if we assume that individuals satisfy Slutsky symmetry, this exactly identifies the Slutsky terms.
\begin{thm}\label{thm:secondmoment}
If individuals obey Slutsky symmetry, the first two moments of demand identify the Slutsky matrix $\mathbb{E}\left[\frac{\partial}{\partial p} \bh^\omega(\bp, u)\right]$.
\end{thm}
\begin{proof}
Using the definition of the conditional moments, we know that 
\begin{equation*}
    \begin{split}
         \frac{\partial}{\partial y}\bM_2(\bp,y) &=  \frac{\partial}{\partial y}\left(\int \bq^\omega(\bp,y)\bq^\omega(\bp,y)^\intercal dF(\omega)\right) \\
         &= \int  \frac{\partial}{\partial y}[\bq^\omega(\bp,y)\bq^\omega(\bp,y)^\intercal] dF(\omega) \\
        &= \int \left[ \frac{\partial}{\partial y}\bq^\omega(\bp,y)\bq^\omega(\bp,y)^\intercal+ \bq^\omega(\bp,y) \left(\frac{\partial}{\partial y}\bq^\omega(\bp,y)\right)^\intercal\right]dF(\omega),
    \end{split}
\end{equation*}
where the second equality follows from interchanging the derivative and integral operators and the third equality from the chain rule. Using the Slutsky equation \eqref{eq:slutsky}, we have that 
\begin{equation*}
    \frac{\partial}{\partial p} \bh^\omega(\bp, u) = \frac{\partial}{\partial p} \bq^\omega(\bp,y) +  \frac{\partial}{\partial y}\bq^\omega(\bp,y) \bq^\omega(\bp,y)^\intercal,
\end{equation*}
which is symmetric due to Slutsky symmetry. Adding this equation to its transpose fetches us
\begin{equation*}
    \begin{split}
        2\frac{\partial}{\partial p} \bh^\omega(\bp, u)&=\frac{\partial}{\partial p} \bq^\omega (\bp,y) + \left(\frac{\partial}{\partial p} \bq^\omega (\bp,y)\right)^\intercal \\
        &\qquad + \frac{\partial}{\partial y}\bq^\omega(\bp,y)\bq^\omega(\bp,y)^\intercal + \bq^\omega(\bp,y) \left(\frac{\partial}{\partial y}\bq^\omega(\bp,y)\right)^\intercal,
    \end{split}
\end{equation*}
such that
\begin{equation*}
    \begin{split}
        \mathbb{E}\left[\frac{\partial}{\partial p} \bh^\omega(\bp, u)\right] & =\frac{1}{2}\left[\frac{\partial}{\partial p}\bM_1(\bp,y)+ \left(\frac{\partial}{\partial p}\bM_1(\bp,y)\right)^\intercal+ \frac{\partial}{\partial y}\bM_2(\bp,y)\right].
    \end{split}
\end{equation*}
\end{proof}

Theorem~\ref{thm:secondmoment} shows that two symmetric models that generate the same conditional mean and variance of demand have the same average substitution. Therefore, the first two moments pin down average substitution under Slutsky symmetry. 

\begin{rem}
    Proposition~\ref{prop:symmetry} and Theorem~\ref{thm:secondmoment} imply that Slutsky symmetry is untestable from the first two moments of demand. That can be seen because there are several values of $\mathbb{E}\left[ \frac{\partial}{\partial y}\bq^\omega(\bp,y) \bq^\omega(\bp,y)^\intercal\right]$ which agree with a mean-variance system, but only one value which arises from a symmetric system. Therefore there must be asymmetric systems which agree with the mean-variance data, and there must also be symmetric systems as we have shown above. This renders symmetry untestable.
\end{rem}

\paragraph{Observable restrictions.} 

Assuming Slutsky symmetry, one can test negative semi-definiteness of the population based on the moments of demand. For instance, using the first two moments, the matrix
    \begin{equation*}
        \bP(\bp,y) = \frac{\partial}{\partial p} \bM_1(\bp,y) + \frac{1}{2} \frac{\partial}{\partial y}\bM_2(\bp,y)
    \end{equation*}
    must be negative semidefinite. This follows from the fact that $\frac{\partial}{\partial p} \bh^\omega(\bp, u)$ is NSD and $\bP(\bp,y)+\bP(\bp,y)^\intercal= 2 \frac{\partial}{\partial p} \bh^\omega(\bp, u)$.\footnote{Note that for a square matrix $\mathbf{A}$ it holds that $\mathbf{v}^\intercal (\mathbf{A} + \mathbf{A}^\intercal) \mathbf{v} = 2\mathbf{v}^\intercal \mathbf{A} \mathbf{v}$.}

Akin to the two-good case, we have similar restrictions on the higher moments of demand. The difference is that the monomial translation for any moment is now a tensor form. The following theorem provides necessary conditions for the moments to be generated by a demand system.
\begin{thm}\label{thm:ratmulti}
For any $n\in \mathbb{N}$, the following $n+1$ tensor form is negative semidefinite:\footnote{We say a tensor form $\mathbf{T}^\omega_n$ is \emph{negative semidefinite} if
\begin{equation*}
    \mathbf{T}^\omega_n(\underbrace{\mathbf{v} \times \mathbf{v} \times \dots \times \mathbf{v}}_\text{$n$ times}) = \sum_{i_1, i_2, \dots, i_n = 1}^{l-1} t^\omega_{i_1, i_2, \dots, i_n} v_{i_1}v_{i_2}\dots v_{i_n} \leq 0, \quad \forall \mathbf{v} \in \mathbb{R}^{l-1}.
\end{equation*}} \footnote{Notice that the form is of order $n+1$, because differentiating a $n$ form with respect to price increases the order of the form.}
\begin{equation*}
    n^{-1} \frac{\partial}{\partial p} \bM_n + (n+1)^{-1}  \frac{\partial}{\partial y}\bM_{n+1}.
\end{equation*}
\end{thm}
\begin{proof}
    The proof is similar to the that of the two-good case. Notice that 
\begin{equation*}
    n^{-1} \frac{\partial}{\partial p} \bM_n + (n+1)^{-1}  \frac{\partial}{\partial y}\bM_{n+1}
\end{equation*}
is the same as the symmetrized sum 
\[\left(\bigotimes_{k=1}^{n-i-1}\bq^\omega(\bp,y)\right)(**)\left(\frac{\partial}{\partial p} \bM_1 + (n+1)^{-1}  \frac{\partial}{\partial y}\bM_{2}\right),\]
where the second term is NSD by the Slutsky equation. Because the product of a NSD matrix and any other tensor form must be NSD, so must the above expression.
\end{proof}

\begin{rem}
    Because the restriction in Theorem~\ref{thm:ratmulti} is a test of negative semidefiniteness (and not of symmetry), any small perturbation of a finite and rationalizable moment sequence is itself also rationalizable. This is because negative semidefiniteness is an open condition.
\end{rem}

\begin{rem}
    Finally, the restrictions in Theorems~\ref{thm:rationalizability} and \ref{thm:ratmulti} do not depend on the levels of the moments, but only on their changes with respect to prices and income. This leads to two fundamental properties of these restrictions. First, if there is additively separable i.i.d. measurement error in the observed demands, these restrictions can still be estimated consistently. Second, none of our restrictions depend on statistical constraints on moments, such as non-negativity (for even moments) or Chebyshev-type tail inequalities.
\end{rem}

\section{Comparison with the quantile-based approach}\label{sec:numexample}
In this section, we compare our moments-based approach with that of the quantile-based approach in the setting with two goods. We first discuss the theoretical connection between both approaches; we then assess their relative power in detecting violations from rationality by means of a numerical example.

\subsection{Theoretical connection}
Similar as before, let $F(q) = \Pr[q^\omega(p,y) \leq q \mid p,y]$ be the conditional CDF of demand and $Q(\tau) = \inf \{q : \tau \leq F(q)\}$ its associated conditional quantile function.\footnote{For notational brevity, we suppress the conditioning on the budget $(p,y)$.} \citet{Dettehoderlein2016} and \citet{hausman2016individual} show that the population of consumers is rational if and only if the inequality
\begin{equation}\label{eq:quantile}
    \begin{split}
        R_Q(\tau) = \frac{\partial Q(\tau)}{\partial p} + \frac{\partial Q(\tau)}{\partial y} Q(\tau) \leq 0,\\
    \end{split}
\end{equation}
holds at every quantile $\tau \in (0,1)$ and budget $(p,y)$.

Our moments-based restrictions implicitly weight these quantile-based restrictions. To see this, notice that 
\begin{equation*}
    \begin{split}
        \Gamma_{n} &= \mathbb{E}\left[\left(\frac{\partial q^\omega(p,y)}{\partial p} - \frac{\partial q^\omega(p,y)}{\partial y} q^\omega(p,y)\right)q^\omega(p,y)^{n}\right]\\
        & =\mathbb{E}\left[\mathbb{E}\left[\frac{\partial q^\omega(p,y)}{\partial p} - \frac{\partial q^\omega(p,y)}{\partial y} q^\omega(p,y) \mid q^\omega(p,y)\right]q^\omega(p,y)^{n}\right] \\
        &=\mathbb{E}\left[R_Q(F(q^\omega(p,y)))q^\omega(p,y)^{n}\right] \\
        &= \int  R_Q(F(q)) q^{n} dF(q) \\
        &= \int_0^1 R_Q(\tau) Q(\tau)^{n} d\tau,
    \end{split}
\end{equation*}
where the second equality follows from the law of iterated expectations. The $n$-th monomial translation is therefore a weighted average across the quantile restrictions $R_Q(\tau)$ with weights $Q(\tau)^{n}
$. For example, the first restriction (i.e., $n=0$) gives equal weight to all quantiles, whereas the second restriction (i.e., $n=1$) gives more weight to the upper quantiles.

\subsection{Numerical example}\label{sec:subnumexample}
 It turns out that even the first three moments of demand already provide considerable power in detecting deviations from rationality.

\paragraph{Setup.} Let individual demand take the form of the linear random coefficients specification
\begin{equation*}
    q^\omega(p,y) = 1- \omega_p p + \omega_y y,
\end{equation*}
where the coefficients that capture unobserved heterogeneity are distributed independently of each other and of prices and income: i.e., $f(\omega_p, \omega_y \mid p, y) = f(\omega_p)f(\omega_y)$.\footnote{Note that in a model with additively separable unobserved heterogeneity, the restriction based on the first two moments of demand provides a necessary and sufficient test for rationality. We therefore focus on the more challenging case where heterogeneity enters multiplicatively.} Both coefficients $\omega_p$ and $\omega_y$ are drawn from uniform distributions with supports $[a_p,b_p]$ and $[a_y, b_y]$, respectively. Demand is rational for consumer $\omega$ at budget $(p,y)$ if the Slutsky inequality $-\omega_p + \omega_y q^\omega(p,y) \leq 0$ holds. This condition is violated, for example, when $\omega_p$ is sufficiently small and $\omega_y$ is sufficiently large.

First consider the restrictions imposed by rationality on the moments of demand. Using the results from Section~\ref{sec:momrational}, for the first three moments of demand we obtain the inequalities
\begin{equation*}
    \begin{split}
        \frac{\partial M_1(p,y)}{\partial p} - \frac{1}{2}\frac{\partial M_2(p,y)}{\partial y} & \leq 0, \\
         \frac{1}{2}\frac{\partial M_2(p,y)}{\partial p} - \frac{1}{3}\frac{\partial M_3(p,y)}{\partial y} &\leq 0, \\
    \end{split}
\end{equation*}
and 
\begin{equation*}
    \begin{split}
        - q_{min}(p,y) \left(\frac{\partial M_1(p,y)}{\partial p} - \frac{1}{2}\frac{\partial M_2(p,y)}{\partial y} \right) + \left(\frac{1}{2}\frac{\partial M_2(p,y)}{\partial p} - \frac{1}{3}\frac{\partial M_3(p,y)}{\partial y}\right)&\leq 0, \\
        q_{max}(p,y) \left(\frac{\partial M_1(p,y)}{\partial p} - \frac{1}{2}\frac{\partial M_2(p,y)}{\partial y} \right) - \left(\frac{1}{2}\frac{\partial M_2(p,y)}{\partial p} - \frac{1}{3}\frac{\partial M_3(p,y)}{\partial y}\right) &\leq 0,
    \end{split}
\end{equation*}
which should hold at every budget $(p,y)$. Notice that $q_{min}(p,y) = 1-b_p p + a_y y$ and $q_{max}(p,y) = 1-a_p p + b_y y$. 

Now consider the restrictions on the quantiles of demand. We will test the quantile restriction in Expression~\eqref{eq:quantile} at $\tau \in \{0.33, 0.50, 0.66\}$. We refer to Appendix~\ref{app:numexample} for closed-form expressions for the moments and quantiles of demand and their associated restrictions.

\begin{figure}[h]
\centering
   \includegraphics[width=1\linewidth]{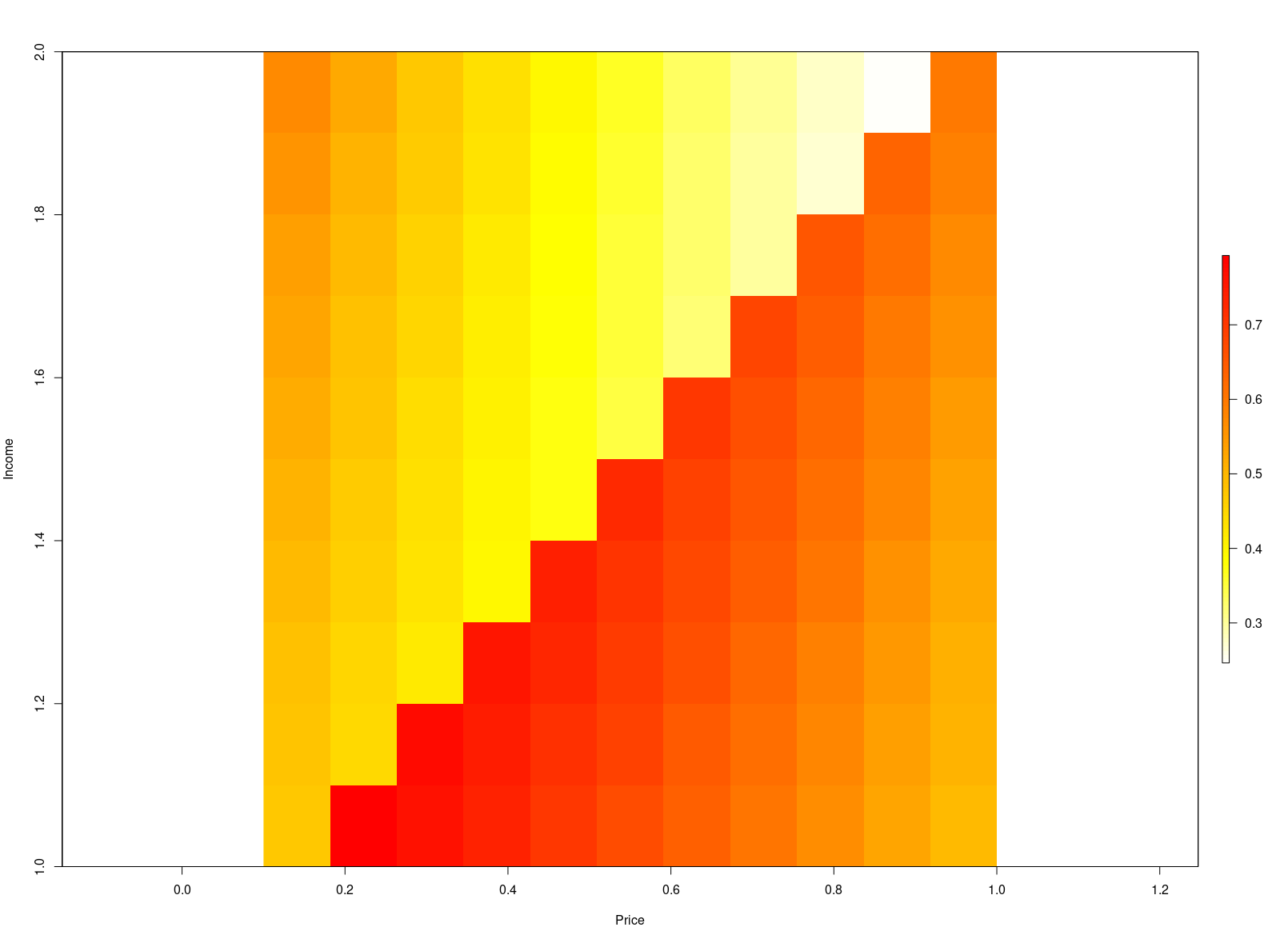}
   \caption{Share of the population that is irrational at budget $(p,y)$}
   \label{fig:frac} 
\end{figure}

\paragraph{Analysis.} We now assess the relative power of the testable restrictions based on the moments and quantiles of demand. To do so, the supports of the random coefficients on price and income are fixed at $\left[a_p, b_p\right] = \left[\frac{1}{3}, 1\right]$ and $\left[a_y, b_y\right] = \left[\frac{1}{3}, \frac{2}{3}\right]$, respectively. We test the theoretical restrictions for all budgets $(p, y) \in \left[\frac{1}{10}, 1 \right] \times [1,2]$. As illustrated by Figure~\ref{fig:frac}, the share of irrational consumers ranges from below 30\% (about the upper right corner) to above 70\% (about the lower left corner) across this space of budgets. 

\begin{figure}[p]
\centering
\begin{subfigure}{0.95\textwidth}
   \includegraphics[width=1\linewidth]{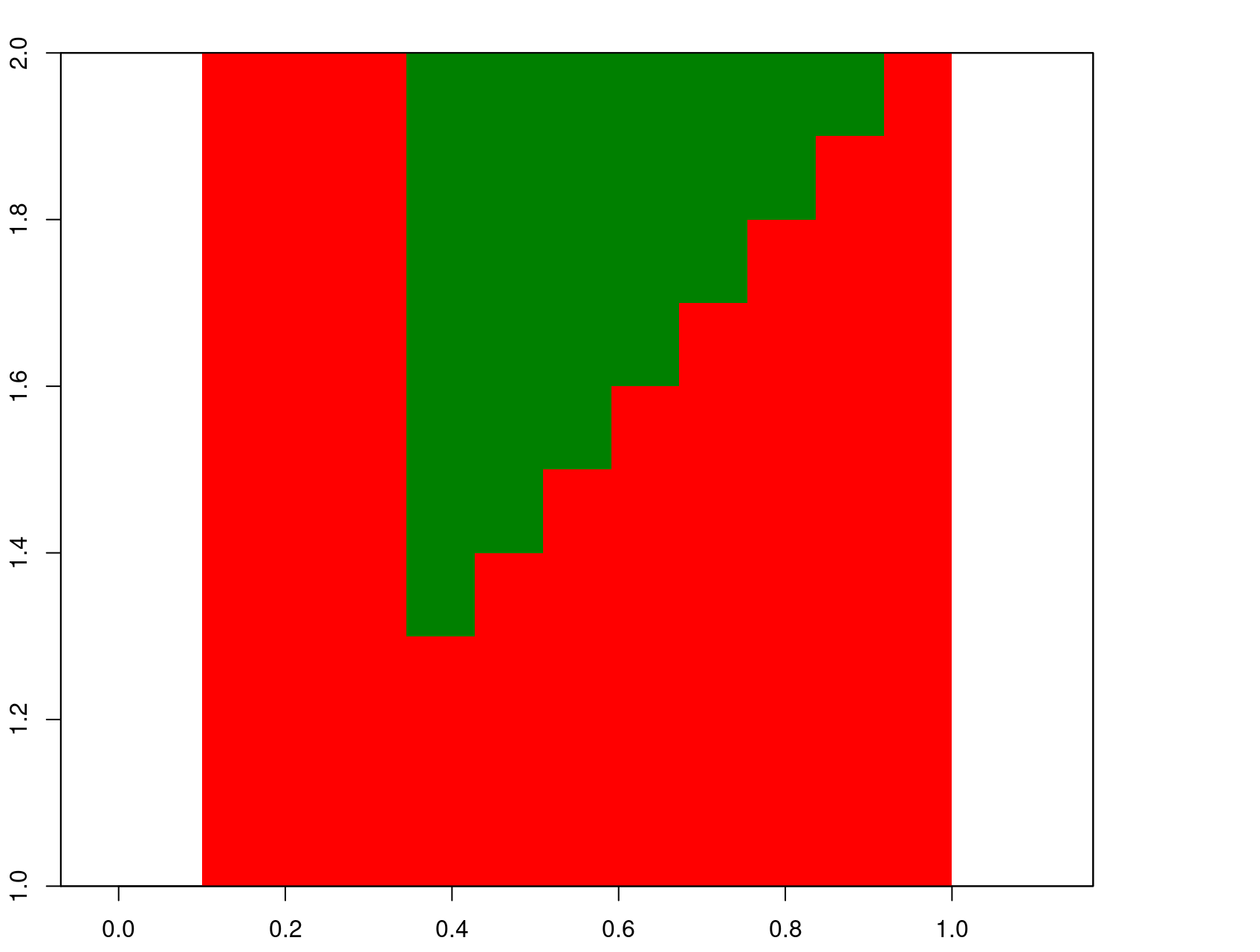}
   \caption{Violations using the first three moments}
   \label{fig:ratmom} 
\end{subfigure}
\hfill
\begin{subfigure}{0.95\textwidth}
   \includegraphics[width=1\linewidth]{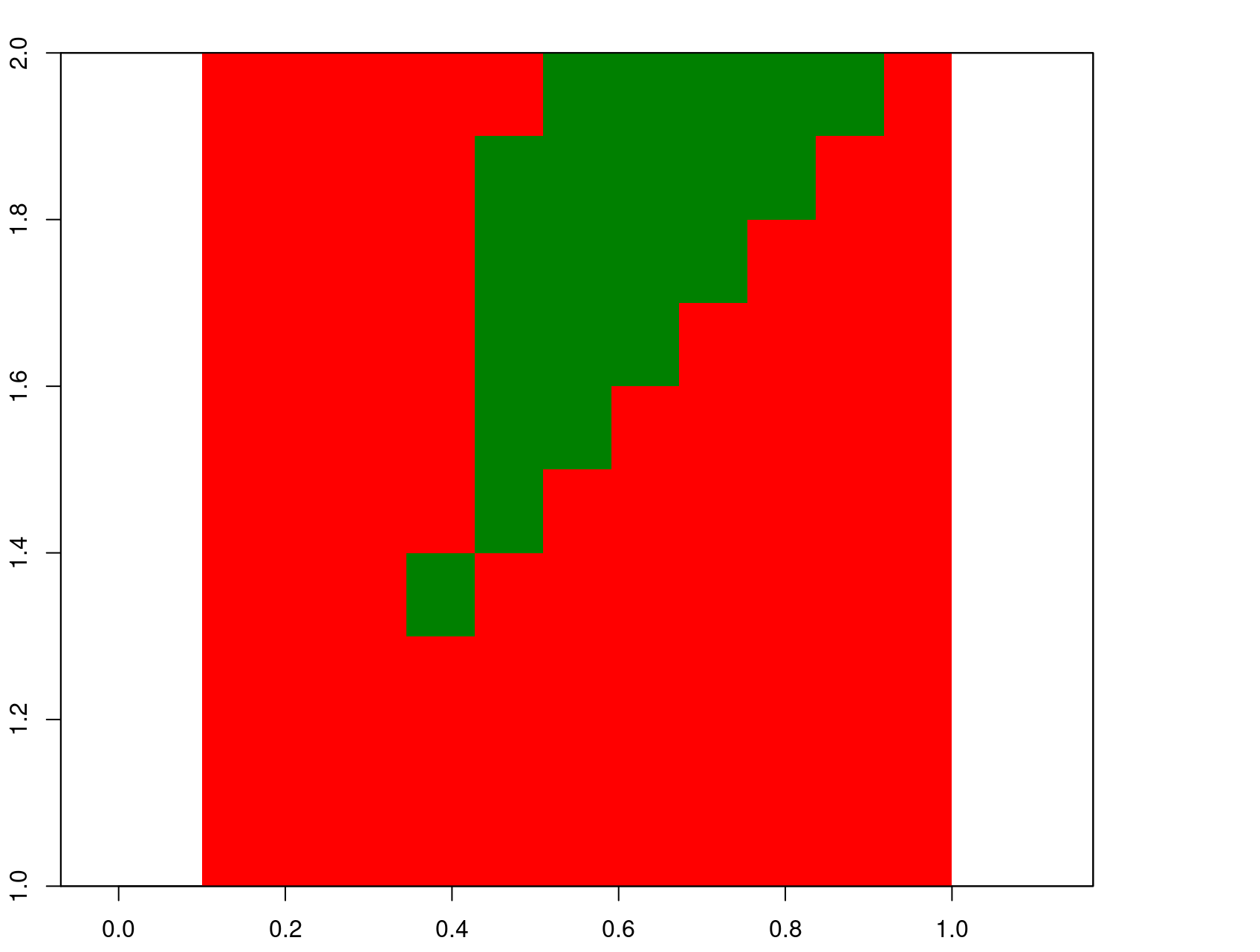}
    \caption{Violations using the quantiles $\tau \in \{0.33, 0.50, 0.66\}$}
    \label{fig:ratquant}
\end{subfigure}
\caption{Tests of rationality based on observable restrictions at budget $(p,y)$}
\label{fig:rat}
\end{figure}

We now present the results of the moment- and quantile-based tests for the same grid of budget sets. In our simple setup, Figure~\ref{fig:ratmom} shows that rationality is rejected based on the first three moments of demand when more than 40 to 50\% of the population of consumers is irrational (indicated in red). This performance is similar to that of the quantile-based approach with quantiles $\tau \in \{0.33, 0.50, 0.66\}$, as illustrated by Figure~\ref{fig:ratquant}. Together, these figures suggest that even in the two-good case, our moment-based approach can deliver a competitive and practical test for consumer rationality.

Moreover, as shown in Appendix~\ref{app:numexample}, there can be substantial variation in the power of the quantile-based tests across different quantiles. In our setup, it is mainly the $\tau = 0.66$ quantile that contributes to the detection of irrationality; by contrast, the $\tau = 0.33$ quantile detects very little violations.\footnote{Notice that in our setup, there are more irrational consumers among those with high amounts of consumption, conditional on prices and income. Therefore, the higher quantiles contribute more to the detection of irrationality.} Moment-based tests, by their averaging nature, are less susceptible to such fluctuations in power.

\section{Applications}\label{sec:applications}

\subsection{Improved precision of demand and welfare estimates}
The restrictions developed in the previous section can be used to improve the precision of estimates of the moments of demand. This is of considerable practical importance: it has recently been shown that knowledge on these moments delivers best approximations to welfare changes using cross-sectional data \citep{maesmalhotra2023robust}. We advance an empirical Bayes approach in the spirit of \citet{fesslerkasy} to shrink the unconstrained estimates towards the theoretical restrictions, if the latter are not rejected by the data.\footnote{This approach differs from those taken by, e.g., \citet*{blundellhorowitzparey}, where the theoretical restrictions are fully imposed on demand estimates. Nevertheless, our restrictions can also be used to discipline nonparametric kernel and series estimators in such a setup.}

The approach hinges on the availability of consistent, albeit potentially noisy, estimators of the moments of demand. For instance, unconstrained estimates can be obtained through local linear regression around a budget $(p, y)$:
\begin{equation*}
    \begin{split}
        (\widehat{\alpha}_n, \widehat{\beta}_{np}, \widehat{\beta}_{ny}) = {\arg \min}_{a_n,b_{np}, b_{ny}} \sum_{i} K_h(p^i - p, y^i-y) \left[(q^i)^n - a_n - b_{np} (p^i - p) - b_{ny} (y^i-y)\right]^2,
    \end{split}
\end{equation*}
where $K_h$ is a kernel function with bandwidth $h$. An estimate $\widehat{\mathbf{V}}$ of the joint variance-covariance matrix of $\{ \widehat{\beta}_{np}, \widehat{\beta}_{ny}\}_{n \in \mathbb{N}}$ can be obtained using the bootstrap.\footnote{Note that only the slope parameters play a role in the subsequent analysis.} Although this approach delivers consistent estimates, we might construct more efficient estimators that incorporate the theoretical restrictions implied by rationality. From Proposition~\ref{prop:res2mom}, we know that the slope parameters of the moments of demand must satisfy a set of linear inequalities. In particular, admissible parameters must lie in the set
\begin{equation*}
    \mathcal{B}_0 = \left\{\mathbf{b}: \frac{1}{n+1} b_{(n+1)p} + \frac{1}{n+2}  b_{(n+2)y}\leq 0, \qquad  n \in \mathbb{N}\right\}.
\end{equation*}

Following \citet{fesslerkasy}, the empirical Bayes approach shrinks the unconstrained slope estimates $\widehat{\boldsymbol{\beta}}$ towards the constrained slope estimates 
\begin{equation*}
    \widehat{\boldsymbol{\beta}}_0(\tau^2) = {\arg \min}_{\mathbf{b}_0 \in \mathcal{B}_0} (\widehat{\boldsymbol{\beta}} - \mathbf{b}_0)^\intercal (\tau^2 \mathbf{I} + \widehat{\mathbf{V}})^{-1} (\widehat{\boldsymbol{\beta}} - \mathbf{b}_0), \\
\end{equation*}
through the weighted average
\begin{equation*}
    \begin{split}
        \widehat{\boldsymbol{\beta}}_{EB}(\tau^2) &= \widehat{\boldsymbol{\beta}}_0(\tau^2) + \left(\mathbf{I} + \frac{1}{\tau^2 }\widehat{\mathbf{V}}\right)^{-1} (\widehat{\boldsymbol{\beta}} - \widehat{\boldsymbol{\beta}}_0 (\tau^2)). 
    \end{split}
\end{equation*}
Intuitively, the nuisance parameter $\tau^2$ captures whether the theoretical restrictions are plausible given the data at hand. When $\tau^2 \rightarrow \infty$, the data strongly reject the theoretical restrictions and the empirical Bayes estimate simply equals the unconstrained estimate: i.e., $\widehat{\boldsymbol{\beta}}_{EB}(\tau^2) = \widehat{\boldsymbol{\beta}}$. On the other hand, when $\tau^2 \rightarrow 0$, the data strongly support the theoretical restrictions and the empirical Bayes estimate equals the constrained estimate: i.e.  $\widehat{\boldsymbol{\beta}}_{EB}(\tau^2) = \widehat{\boldsymbol{\beta}}_0(\tau^2)$. The nuisance parameter can be estimated in a data-driven fashion through the minimization of Stein's unbiased risk estimate:
\begin{equation*}
    \begin{split}
        \widehat{\tau}^2 = \arg \min || \widehat{\boldsymbol{\beta}} - \widehat{\boldsymbol{\beta}}_{EB}(\tau^2) ||^2 + 2~\text{trace}\left(\nabla (\widehat{\boldsymbol{\beta}} - \widehat{\boldsymbol{\beta}}_{EB}(\tau^2)) \widehat{\mathbf{V}}\right).
    \end{split}
\end{equation*}
A feasible empirical Bayes estimate then follows directly from the plug-in principle:
\begin{equation*}
    \widehat{\boldsymbol{\beta}}_{EB} = \widehat{\boldsymbol{\beta}}_{EB}(\widehat{\tau}^2).
\end{equation*}

\subsection{Existence of a normative representative consumer}
Exact aggregation of consumer demands is a long studied economics problem.\footnote{For foundational work on exact aggregation, see \citet{gorman53,nataf1954questions, antonelli71}.} Aggregation fundamentally boils down to two questions:
\begin{enumerate}[label=(\roman*)]
    \item When can an aggregate demand be rationalized by a single preference relation? That is, when does there exist a \emph{positive} representative consumer?\footnote{There exists a positive representative consumer when average demand satisfies the Slutsky conditions: i.e., $\frac{\partial \bM_1(\bp,y)}{\partial \bp}+\frac{\partial \bM_1(\bp,y)}{\partial y}\bM_1(\bp,y)^\intercal$ is symmetric and NSD at every budget $(\bp, y)$.}
    \item When is this preference relation is welfare relevant? That is, when does there exist a \emph{normative} representative consumer?
\end{enumerate}

\citet*{gorman53} showed that for (ii) to be true for arbitrary distributions of income, income effects must be constant across the population and independent of income. In other words, these \emph{Gorman conditions} require consumers' income expansion paths to be linear and parallel. \citet*{jerison1994optimal, jerison96} show that the conditions for the existence of a positive normative consumer are much weaker than the Gorman conditions. Nevertheless, this positive representative consumer may not be welfare-relevant for society, violating (ii). For instance, \citet*{dow1988consistency} show that this agent may be Pareto inconsistent, preferring situations in which each agent of society is worse off. 

Clearly if only the first moment of demand is observed, one cannot distinguish between normative and positive representative agents. 
However, when mean-variance data is available, we can test for aggregation conditions.
We suppose a representative agent satisfies normativity.
\begin{mydef}\label{ass:cond}
    A positive representative satisfies \emph{normativity} if
    \begin{equation*}
    \mathbb{E}\left[\frac{\partial \bh^\omega (\bp, u)}{\partial \bp}\right]=\frac{\partial \bM_1(\bp, y)}{\partial \bp}+ \bM_1(\bp, y)\frac{\partial \bM_1(\bp, y)}{\partial y}^\intercal,
    \end{equation*}
    at a budget $(\bp, y)$.
\end{mydef}
Normativity requires the representative agent to approximate average welfare measures for society. If this condition holds, Hicksian demands aggregate locally. It is equivalent to requiring $\Cov\left(\frac{\partial \bq^\omega (\bp,y)}{\partial y}, \bq^\omega(\bp,y)^\intercal\right) = \mathbf{0}$. The following proposition provides a test of normativity.
\begin{prop}\label{thm:normative}
Suppose a population of rational consumers generates a positive representative agent. This representative agent satisfies normativity if and only if 
\begin{equation*}
    \frac{\partial}{\partial y}\left(\bM_2(\bp, y)-\bM_1(\bp, y)\bM_1(\bp, y)^\intercal\right)=\mathbf{0},
\end{equation*}
at all budgets $(\bp,y)$.
\end{prop}

\begin{proof}
For the $\Longrightarrow$ part, notice that normativity implies that
\begin{align*}
    & \frac{\partial}{\partial y}\left[\bM_2(\bp, y)-\bM_1(\bp, y)\bM_1(\bp, y)^\intercal\right]\\
     &\quad = \mathbb{E} \left[ \frac{\partial}{\partial y}\bq^\omega(\bp,y)\bq^\omega(\bp,y)^\intercal\right] + \mathbb{E}\left[\bq^\omega(\bp,y) \left(\frac{\partial}{\partial y}\bq^\omega(\bp,y)\right)^\intercal\right] \\
    &\qquad - \mathbb{E} \left[ \frac{\partial}{\partial y}\bq^\omega(\bp,y)\right] \mathbb{E}\left[\bq^\omega(\bp,y)^\intercal\right] + \mathbb{E}\left[\bq^\omega(\bp,y) \right] \mathbb{E}\left[\left(\frac{\partial}{\partial y}\bq^\omega(\bp,y)\right)^\intercal\right] \\
    &\quad= \mathbf{0},
\end{align*}
since $\Cov\left(\frac{\partial \bq^\omega (\bp,y)}{\partial y}, \bq^\omega(\bp,y)^\intercal\right) = \mathbf{0}$.

For the $\Longleftarrow$ part, observe that 
\[\frac{\partial}{\partial y}\left(\bM_2(\bp, y)-\bM_1(\bp, y) \bM_1(\bp, y)^\intercal\right) = \mathbf{0}  \implies  \frac{\partial }{\partial y} \bM_2(\bp, y)= \frac{\partial }{\partial y}(\bM_1(\bp, y) \bM_1(\bp, y)^\intercal).\]
From the proof of Theorem~\ref{thm:secondmoment}, we know that
\begin{align*}
\mathbb{E}\left[\frac{\partial}{\partial \bp} \bh^\omega(\bp, u)\right] & =\frac{1}{2}\left[\frac{\partial}{\partial \bp}\bM_1(\bp,y)+ \left(\frac{\partial}{\partial \bp}\bM_1(\bp,y)\right)^\intercal+ \frac{\partial}{\partial y}\bM_2(\bp,y)\right]\\
        &=\frac{1}{2}\left[\frac{\partial}{\partial \bp}\bM_1(\bp,y)+ \left(\frac{\partial }{\partial \bp}\bM_1(\bp,y)\right)^\intercal+ \frac{\partial }{\partial y}(\bM_1(\bp,y) \bM_1(\bp,y)^\intercal)\right]
        \\
        &=\frac{1}{2}\left[\frac{\partial}{\partial \bp}\bM_1(\bp,y)+ \frac{\partial }{\partial y}\bM_1(\bp,y)\bM_1(\bp,y)^\intercal\right] \\
        & \quad +\frac{1}{2}\left[\left(\frac{\partial }{\partial \bp}\bM_1(\bp,y)\right)^\intercal+ \bM_1(\bp, y) \frac{\partial }{\partial y}\bM_1(\bp,y)^\intercal\right]^\intercal.
\end{align*}
Both terms are equal as the Slutsky matrix of the positive representative agent is symmetric by assumption.
\end{proof}

\begin{rem}
    Note that the condition we have put forward is weaker than that of Gorman. The Gorman condition can be written formally as \begin{equation*}\label{eq:gormann-equal}
    \quad\frac{ \partial \bq^\omega(\bp,y)}{ \partial y}=\frac{\partial \bq^{\omega'}(\bp,y)}{\partial  y}= \mathbf{c}, \qquad \forall (\omega, \omega'), \forall (\bp, y),
    \end{equation*}
    and implies that $\Cov\left(\frac{\partial \bq^\omega (\bp,y)}{\partial y}, \bq^\omega(\bp,y)^\intercal\right) = \mathbf{0}$. This guarantees the existence of a normative representative consumer per Proposition~\ref{thm:normative}.

    However, the reverse implication is not true. The existence of a normative representative consumer, and therefore $\Cov\left(\frac{\partial \bq^\omega (\bp,y)}{\partial y}, \bq^\omega(\bp,y)^\intercal\right) = \mathbf{0}$, does not necessarily imply the Gorman condition. 
\end{rem}

\begin{rem}
    We can think of the term $\frac{\partial}{\partial y}\left(\bM_2(\bp, y)-\bM_1(\bp, y)\bM_1(\bp, y)^\intercal\right)$ as a measure for the ``distance" from normativity or how badly the representative agent mimics average welfare. 
\end{rem}


\section{Conclusion}\label{sec:conclusion}
In this paper, we show that the conditional moments of demand contain empirical content and can be used to test individual rationality, specifically the negative semidefiniteness of the Slutsky matrix. For the common two-good case, we characterize rationality using sequences of the statistical moments of demand.

Considering stochastic rationalizability, it is still a wide-open question as to whether Slutsky symmetry can be tested with cross-sectional data and if so, how to construct tests. Asking if symmetry carries any empirical content at the level of cross-sections would in itself be a very interesting question. We conjecture that Slutsky symmetry is untestable in a cross-section.

We hypothesize that the restrictions developed in this paper can also be used to set-identify parameters in macroeconomic models with heterogeneous agents based on a few coarse aggregate moments.

\newpage
\bibliography{Raghav}

\clearpage
\begin{center}
    {\huge Online Appendix \\
    \vspace{10pt}
    \Large \textit{Moments of Demand and Stochastic Rationalizability}
  }
\end{center}

\appendix

\section{Regularity conditions} \label{app:techcon}
Every individual's demand function $\bq^\omega(\bp,y)$ needs to be infinitely differentiable in $\bp,y$ at all $\bp,y \in \mathcal{P} \times \mathcal{Y}$. This is ensured by the following condition.
\begin{assumption}
    Every individual's preferences are continuous, strictly convex, and locally nonsatiated. The associated utility functions $u^\omega$ are infinitely differentiable everywhere.
\end{assumption}

The following condition ensures that the dominated convergence theorem holds. This allows us to interchange limits and integrals.
\begin{assumption}
    There exists a function $g : \Omega \rightarrow \mathbb{R}$ such that for all $\bp,y \in \mathcal{P} \times \mathcal{Y}$ and $n,m \in \mathbb{N}$ it holds that $\Vert \myvec(D_{p^n, y^m} \bq^\omega(\bp,y))\Vert \leq g(\omega)$ with $\int g(\omega) dF(\omega) < \infty$.
\end{assumption}

Finally, we require that all moments exist and are finite.
\begin{assumption}
    For all $n \in \mathbb{N}$, it holds that
    \begin{equation*}
        \mathbb{E}\left[|\mathbf{T}^\omega_n(\bp,y)|\right] < \infty.P
    \end{equation*}
\end{assumption}

\section{Additional results for Section~\ref{sec:subnumexample}} \label{app:numexample}

\paragraph{Moments of demand.}
Following the setup of Section~\ref{sec:numexample}, the moments of unobserved heterogeneity can be expressed as
\begin{equation*}
    \begin{split}
        \mu_s &= \mathbb{E}[\omega_s] = \frac{a_s + b_s}{2}, \\
        \sigma^2_s &= \mathbb{E}[\omega_s^2] = \frac{a_s^2 + a_s b_s + b_s^2}{3}, \\
        \tau^3_s &=\mathbb{E}[\omega_s^3] = \frac{(a_s + b_s) (a_s^2 + b_s^2)}{4},
    \end{split}
\end{equation*}
for $s \in \{p, y\}$. Straightforward calculations imply that
\begin{equation*}
    \begin{split}
        \frac{\partial M_1(p,y)}{\partial p} - \frac{1}{2}\frac{\partial M_2(p,y)}{\partial y} &= -\mu_p + \mu_y -\mu_p\mu_yp + \sigma_y^2 y, \\
         \frac{1}{2}\frac{\partial M_2(p,y)}{\partial p} - \frac{1}{3}\frac{\partial M_3(p,y)}{\partial y} &= -\mu_p + \mu_y - \mu_p\mu_y (y -2p) +  \sigma_p^2 (p + \mu_yp^2) + 2\sigma_y^2 y (1 -\mu_p p) +
  \tau_y^3 y^2. \\
    \end{split}
\end{equation*}

\paragraph{Quantiles of demand.}
Again following the setup of Section~\ref{sec:numexample}, the conditional CDF of demand can be written as 
\begin{equation*}
    \begin{split}
            F(q \mid p,y) &= \Pr[q^\omega(p,y) \leq q \mid p,y] \\
            &= \Pr[1-\omega_p p + \omega_y y \leq q \mid p,y] \\
            &= \int_{a_p}^{b_p} \int_{a_y}^{\frac{q + \omega_p p - 1}{y}} f(\omega_p, \omega_y \mid p,y) d\omega_p d\omega_y.
    \end{split}
\end{equation*}
Under our distributional assumptions, this CDF has a closed-form solution:
\begin{equation*}
    \begin{split}
        F(q \mid p,y) &= \frac{1}{(b_p - a_p)(b_y - a_y)}\int_{a_p}^{b_p} \int_{a_y}^{\frac{q + \omega_p p - 1}{y}} d\omega_p d\omega_y \\
        &= \begin{cases}
            0, & \text{ if } \frac{q + \omega_p p - 1}{y} \leq a_y, \\
            \frac{1}{(b_y-a_y)y} \left(q - a_y y  + \frac{1}{2}(b_p+a_p)p- 1 \right), & \text{ if } a_y < \frac{q + \omega_p p - 1}{y} < b_y, \\
            1, & \text{ if } \frac{q + \omega_p p - 1}{y} \geq b_y. \\
        \end{cases}
    \end{split}
\end{equation*}
Inverting this CDF gives the quantile demand function
\begin{equation*}
    \begin{split}
        Q(\tau \mid p, y) = ((1-\tau) a_y + \tau b_y)y - \frac{1}{2}(b_p+a_p)p  + 1,
    \end{split}
\end{equation*}
for all $\tau \in (0,1)$. The Slutsky restriction at quantile $\tau$ for the budget set $(p,y)$ therefore takes the form
\begin{equation*}
    \begin{split}
        0 &\geq \frac{\partial Q(\tau \mid p, y)}{\partial p} + \frac{\partial Q(\tau \mid p, y)}{\partial y} Q(\tau \mid p, y) \\
        &= - \frac{1}{2}(b_p+a_p) + ((1-\tau) a_y + \tau b_y) Q(\tau \mid p, y).
    \end{split}
\end{equation*}

\begin{figure}
\centering
\begin{subfigure}{0.58\textwidth}
   \includegraphics[width=1\linewidth]{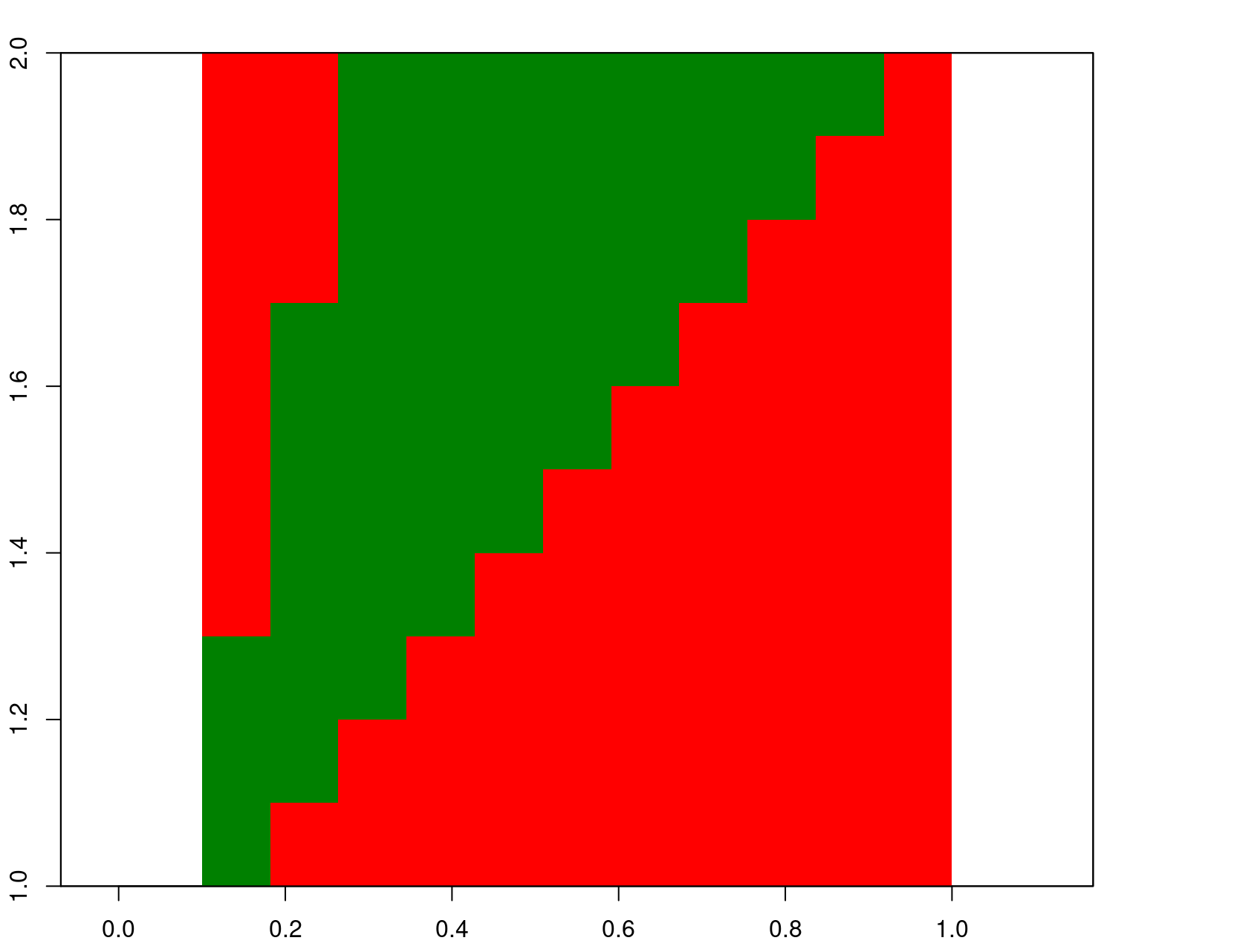}
   \caption{Violations using the first two moments}
   \label{fig:ratmom2} 
\end{subfigure}
\hfill
\begin{subfigure}{0.58\textwidth}
   \includegraphics[width=1\linewidth]{test_mom_all.png}
   \caption{Violations using the first three moments}
    \label{fig:ratmom3}
\end{subfigure}
\caption{Tests of rationality based on observable restrictions at budget $(p,y)$}
\label{fig:ratapp1}
\end{figure}

\paragraph{Analysis.}
Finally, we present some additional results for the numerical example. Figure~\ref{fig:ratapp1} displays the power of the moment-based approach in detecting irrationality using the first two moments (in Figure~\ref{fig:ratmom2}) and the first three moments (in Figure~\ref{fig:ratmom3}) of demand.\footnote{The latter is included for completeness sake; it is identical to that in Figure~\ref{fig:ratmom}.} Although the availability of the third moment provides additional power, the first two moments are already able to detect the most sizeable deviations from rationality. 

\begin{figure}[p]
\centering
\begin{subfigure}{0.58\textwidth}
   \includegraphics[width=1\linewidth]{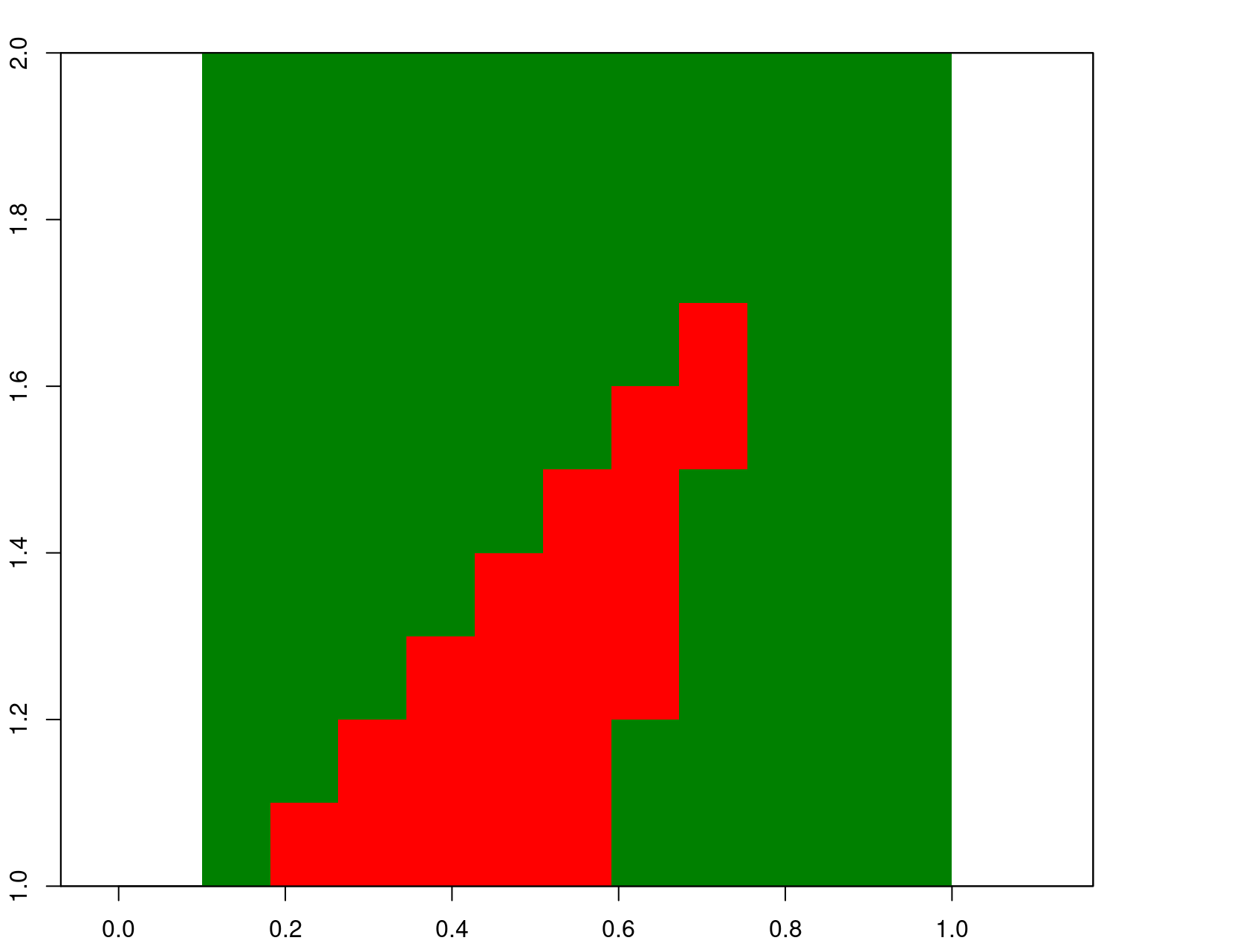}
   \caption{Violations using the quantile $\tau = 0.33$}
   \label{fig:ratquant33} 
\end{subfigure}
\hfill
\begin{subfigure}{0.58\textwidth}
   \includegraphics[width=1\linewidth]{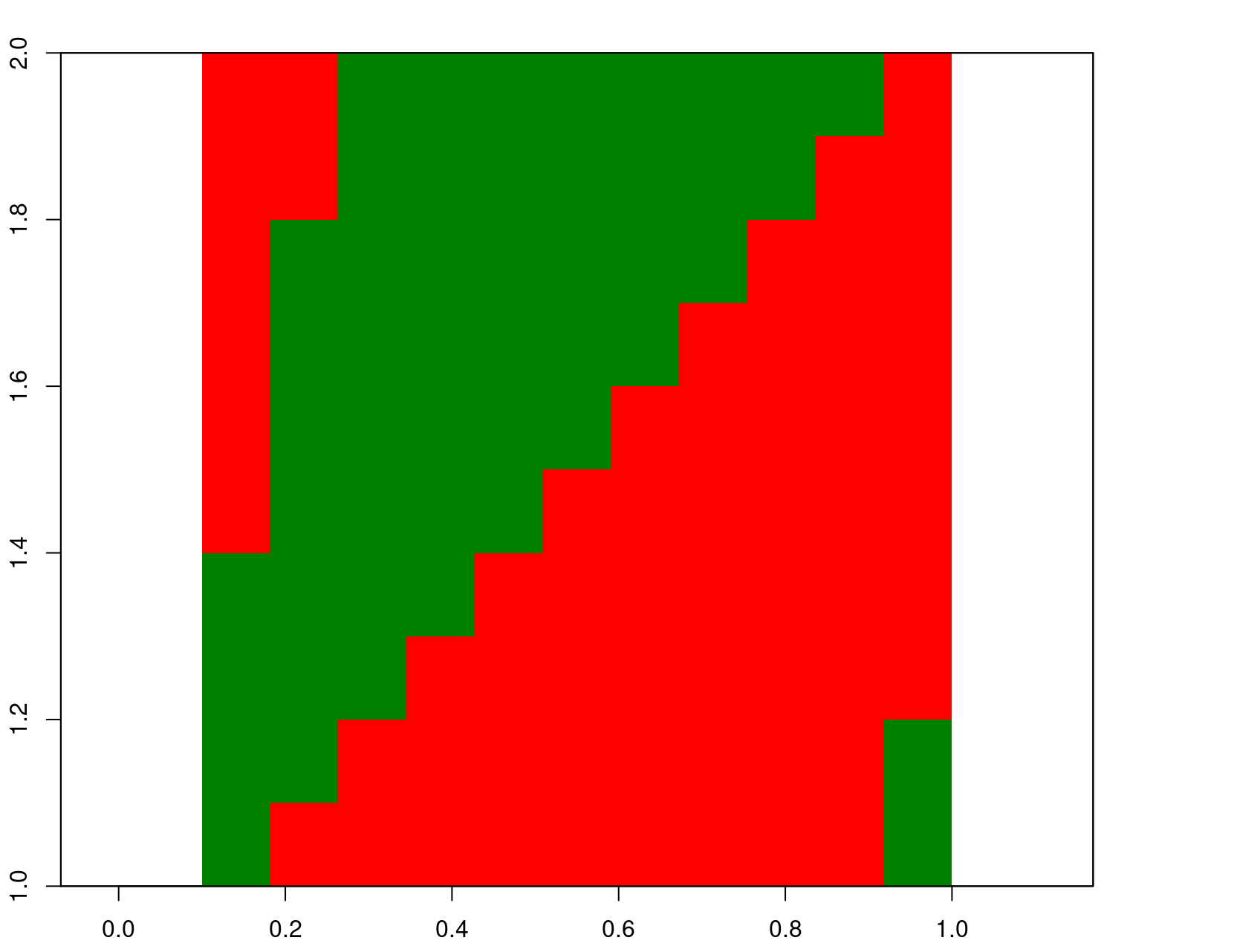}
   \caption{Violations using the quantile $\tau = 0.50$}
    \label{fig:ratquant50}
\end{subfigure}
\hfill
\begin{subfigure}{0.58\textwidth}
   \includegraphics[width=1\linewidth]{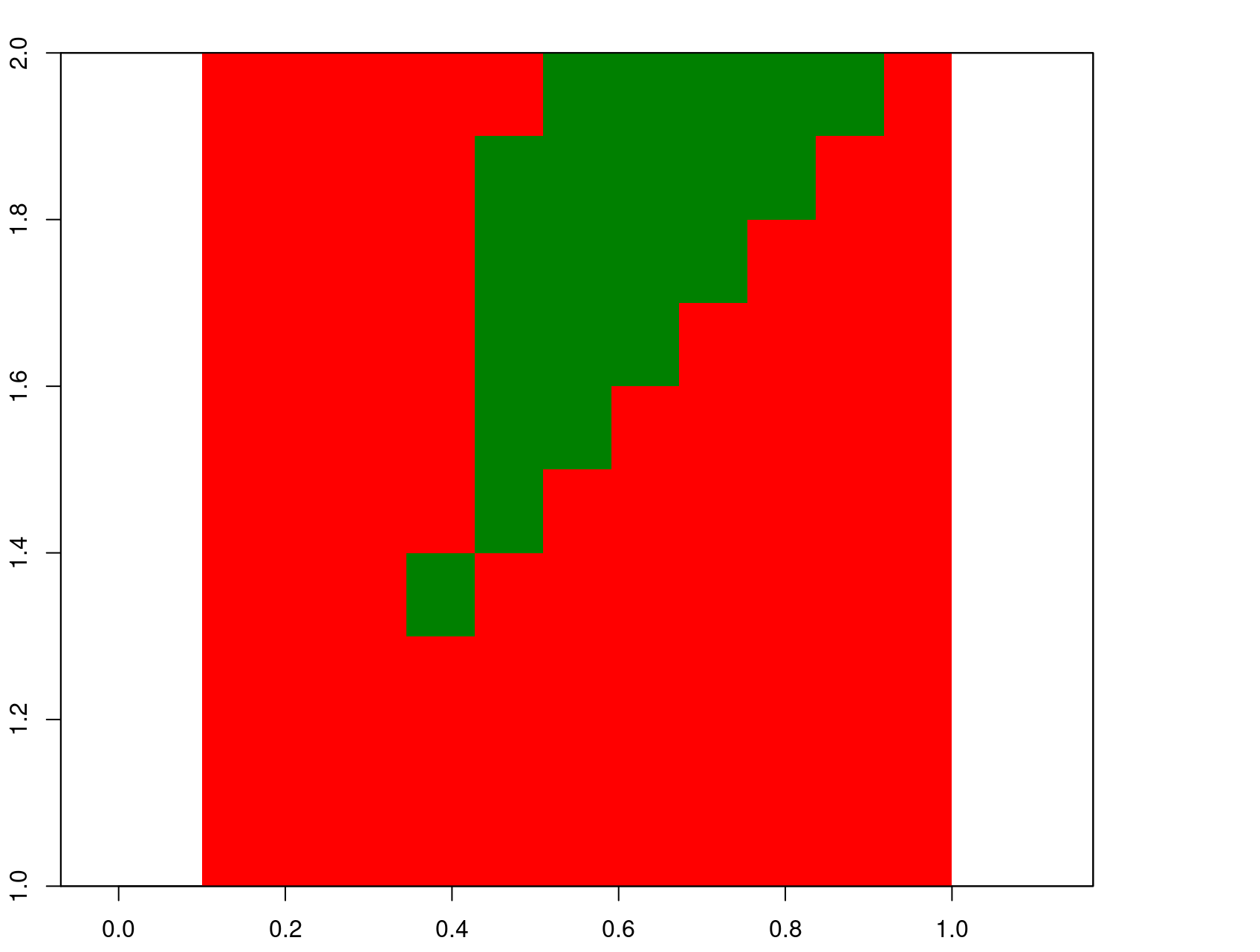}
   \caption{Violations using the quantile $\tau = 0.66$}
    \label{fig:ratquant66}
\end{subfigure}
\caption{Tests of rationality based on observable restrictions at budget $(p,y)$}
\label{fig:ratapp2}
\end{figure}

Figure~\ref{fig:ratapp2} displays the power of the quantile-based approach for the quantiles $\tau \in \{0.33, 0.50, 0.66\}$. As previewed in the main text, the lower quantile contributes little to the detection of irrationality in our setup.

\end{document}